# On the Fundamental Properties of Coupled Oscillating Systems

Danil Doubochinski and Jonathan Tennenbaum

*Quantix - Société de Recherche et Développement en Technique Vibratoire*
*86, Rue de Wattignies, 75012 Paris, France*
danil.doubochinski@gmail.com ; tennenbaum@debitel.net

## Abstract

The present paper presents a new general conception of interaction between physical systems, differing significantly from that of both classical physics and quantum physics as generally understood. We believe this conception could provide the basis for a coherent understanding of several classes of natural phenomena that until now have been studied only in a piece-meal fashion. For example:
1) the universal tendency for physical systems to associate together into stable dynamic formations;
2) the seemingly unlimited capacity for generation of physical objects in Nature, including the emergence of coherent, quantized states in physical systems of the most varied nature;
3) the existence of close couplings between processes whose characteristic length-scales, frequency- and energy-ranges differ by many orders of magnitude.
The proposed conception first emerged in connection with experimental studies of the nonlinear behavior of coupled electromagnetic oscillators, and the discovery of two fundamental phenomena that had been overlooked in the classical theory of oscillations: the quantization of amplitudes as a result of so-called argumental interactions, and the spontaneous aggregative behavior of multiply-coupled resonators placed in a high-frequency field. The essential features of these phenomena are summarized in the first two sections of the paper, after which we demonstrate how the underlying physical principles can be combined under a single notion of interaction, providing a mechanism by which a practically unlimited wealth of physical objects could be generated by the interaction of just a few. The final section puts forward some preliminary ideas about the electromagneto-mechanical dual nature of physical objects as oscillatory processes, suggesting a universal scope for the proposed conceptions.

# Introduction

In the opinion of the authors, one of the main obstacles to advance of natural science today, lies in the lack of an adequate, *unified conception* of the interactions between physical systems. Such a conception would provide the basis for a coherent understanding of the following broad classes of natural phenomena, which until now have been studied only in a piece-meal fashion and from divergent points of view in physics, chemistry, astronomy and biology.

**First**, *the universal tendency for physical systems to associate together to form stable or quasi-stable dynamic formations.*

> This "social" behavior manifests itself at every scale of magnitude, from the level of groups of galaxies, steller associations, the solar system etc. down to the organization of matter on the supramolecular, molecular, atomic and subatomic scales. Although commonly attributed to a variety of different causes, aggregative behavior is in fact a universal tendency of both nonliving and living states of matter.

**Second**, *the seemingly unlimited capacity for generation of physical objects in Nature, including the emergence of coherent, quantized states in physical systems of the most varied nature -- states which constitute distinct physical objects ("particles") in their own right.*

> This second class of phenomena embraces the nature and mechanism of generation of those entities (systems, processes) in the macro-and micro-worlds, which possess a quality of wholeness and distinct individual characteristics, different from those of their component parts, in such a way, that we can justifiably speak of the emergence of a *new physical object.* We believe this phenomenon is more universal, and involves a more fundamental principle of nature, than can be accounted for in the framework of existing physical theory.

**Third**, *the existence of close couplings between processes, whose characteristic length-scales, frequency- and energy-ranges differ by many orders of magnitude.*

> Such couplings are presupposed by the very existence and stability of microscopic objects such as atoms and molecules, by quantum phenomena generally, and by the overall coherence of the world around us. This conclusion is often overlooked due to the habit of thinking, carried over from Newtonian classical physics, which assumes a categorical separation between the interactions of a physical object with "external" objects, on the one hand, and the processes by which that object is constituted and stabilized – processes considered as "internal" to the object –, on the other. We shall discuss this point further in Section 3 below.[1]

In the present paper we shall suggest how the three classes of phenomena, just described, might better be understood from the standpoint of a new, unified conception of physical interactions.

---

[1] An important special case where this issue arises in contemporary physics, is the strong dependence of beta-decay rates of certain radioactive atomic nucleii upon the surrounding electron configurations, demonstrated in a number of experiments. See reference [1].

Interestingly, the new conception first emerged in connection with experimental studies on the behavior of coupled electrical and electromechanical oscillators -- a domain of physics long since believed to have been completed in terms of its essentials, and where fundamental new discoveries were hardly to be expected.

In 1968-69, Danil and Yakov Doubochinski, then students at Moscow University, discovered two basic phenomena which had been overlooked in the classical treatment of coupled oscillating systems.

(1) the phenomenon of so-called argumental interaction between oscillating systems, and the existence of a "Macroscopic Quantum Effect (MQE)" in systems formed by argumentally-interacting oscillators. The principle of argumental interactions will be discussed in Section 1 below.

(2) the tendency of resonators (such as classical LCR circuits), placed in a high-frequency field and free to move in space, to assemble themselves into stable formations, in which fluctuations in the inductive, resistive and capacitive couplings between the resonators permit the coupled system to adapt its absorptive characteristics to the characteristics of the field. We refer to this effect, which will discussed in the Section 2 below, as *the aggregational tendency of coupled oscillating systems.*

Experimental and theoretical investigations, carried out in the subsequent period, indicate that these two effects could be realized at the same time, resulting in a "*Macroscopic Quantum Effect in Systems of Coupled Oscillating Systems (MQECOS)*" having many important theoretical and practical implications. *In particular, we believe it could provide the starting-point for identifying the common physical mechanism for all three of the phenomena of Nature, mentioned above.*

In the following two sections we shall summarize the results of experimental and theoretical investigations of these two basic phenomena, carried out by the Doubochinskis and their collaborators in the 1970s and 1980s, but still not widely known in the scientific community. In the third and fourth sections we elaborate a new conception of physical interactions, growing out of the combination of the two phenomena. In a subsequent paper we shall sketch a new hypothesis on the origin of particles, which is suggested by the MQECOS.

## 1. Argumental interactions and the Macroscopic Quantum Effect (MQE)

The Macroscopic Quantum Effect (MQE) arises when two or more oscillating systems, having widely differing frequencies, become coupled to each other by interactions having a specific phase-dependent character -- so-called *argumental interactions.* Ensembles of argumentally interacting macroscopic oscillators possess a discrete set of stable quasi-stationary modes and other characteristics strikingly similar to microscopic quantum physical objects. The theory and experimental demonstration of the MQE is developed in detail in the scientific literature, and is summarized in the articles [2,3]. For the present purposes, we merely recall the main features of Doubochinski's pendulum, which is the classical experimental demonstration of the MQE, and then extend them to argumentally interacting systems in general.

The argumental pendulum (Figure 1) is composed of two interacting oscillatory processes: (1) a pendulum arm with a natural frequency on the order of 1-2 Hz, with a small permanent magnet fixed at its moving end; and (2) a stationary electromagnet (solenoid) positioned under the equilibrium point of the pendulum's trajectory and supplied with alternating current whose frequency can range from tens to thousands of hertz. The pendulum arm and solenoid are configured in such a way, that the pendulum arm interacts with the oscillating magnetic field of the solenoid only over a limited portion of its trajectory. This *spatial inhomogeniety* of the interaction allows the pendulum to self-regulate its exchange of energy with the magnetic field.

Released from any given position, the pendulum's motion evolves into a stable, very nearly periodic motion whose amplitude takes one of a *discrete array of possible values* (Figure 2). The stability of each amplitude is maintained by a constant adjustment (via fluctuations) of the phase relationship between the pendulum and the high-frequency field. Through its interaction with the field over a given period of oscillation, the pendulum extracts an amount of energy exactly sufficient, on average, to compensate its frictional losses for the same period. The values of the quantized amplitudes -- and the corresponding energies of the quantized modes -- are essentially *independent* of the strength of the alternating current supplied to the electromagnet, over a very large range [2,3].

The theory of the MQE is summarized in our paper [2]. There we show how the *phase-frequency modulation* of the interaction between the electromagnet and the pendulum, produced by the pendulum's own motion, leads to the emergence of a discrete spectrum of stable amplitudes.

The maintenance of these quantized amplitudes depends upon a new mechanism of exchange of energy between oscillating systems, differing *fundamentally* from the mechanisms studied in the classical theory of oscillations, in particular the familiar case of "forced oscillations" of an oscillator under the action of a periodic external force. In the classical case, a significant exchange of energy between the oscillator and the external signal occurs *only* when the frequency of the external force is very close to the proper frequency of the oscillator. In the case of the argumental pendulum, by contrast, the frequency of the magnetic field, from which the pendulum extracts the energy necessary to maintain its undamped motion, can be *two or more orders of magnitude higher* than that of the pendulum arm itself. The latter is close to the proper frequency of the undisturbed pendulum.

The ability of argumental interactions to efficiently convert oscillatory energy over such a large gap of frequencies, is inseparably connected with the role of fluctuations and self-regulating behavior. In the classical (Newtonian) mode of interaction, a system being acted upon by an external force is virtually the "slave" of that force; the Newtonian concept of force leaves no room for a true mutual interaction and mutual adaptation of the interacting systems to each other. The argumental pendulum, by contrast, remains "*its own master*": By shifting the phase of entry into the interaction zone, it can self-regulate its exchange of energy with the electromagnet.

Careful observation and reflection on the behavior of the argumental pendulum results in a number of conclusions of *fundamental importance*.

Firstly, each stable, quasi-periodic mode (amplitude) of the pendulum, resulting from the interaction of the pendulum arm and the magnetic field, constitutes an oscillating system in its own right, with its own distinct energetic and other physical parameters. In such a situation, we can speak of the simultaneous existence of no less than *three* oscillating systems:

System A = the classical oscillator formed by the pendulum arm interacting with the gravitational field of the Earth;
System B = the electromagnet (considered together with its alternating current power source), and System C = the compound oscillating system constituted from the interaction of (A) and (B) in any one of the stable amplitudes.

It is extremely important to observe, that while interacting to form system C, the physical parameters of systems A and B are only slightly modified; each retains its own "identity" and integrity as an oscillating system. Thus, the pendulum arm (System A) continues its basic periodic motion under gravity, while at the same time undergoing fluctuations in amplitude and phase as a result of the action of the magnetic field. The system of the electromagnet and its current source (System B) retains its basic frequency and amplitude characteristics, while at the same time experiencing periodic fluctuations due to currents induced by the motion of the pendulum's permanent magnet through the interaction zone.

The fact, that the two coupled systems A and B retain their essential integrity in the compound system, expresses a very special property of the mechanism of argumental interactions, which distinguishes it fundamentally from the much more rigid classical (and also quantum-mechanical) forms of coupling. In the latter cases, the parameters of the compound systems are not only drastically modified, but the subsystems lose their independence entirely, "melting" into a single combined system with (generally speaking) entirely different parameters. We shall discuss this point further in the following section.

Secondly, as already pointed out, the stable functioning of the coupled system, in a given mode, depends upon the role *phase fluctuations*. Thus, in place of the rigid, "dead" form of Newtonian coupling, argumental interactions have the living, dynamic character which we justifiably attribute to true physical objects. We shall elaborate on our understanding of the term, "physical object" in the third section of this paper.

The above considerations lead to an interesting and rather peculiar form of "arithmetic".

In the classical theory of oscillating systems, the coupling of two oscillators produces a new system – a result we might describe, in symbolic form, as follows:

A + B = C.

Note, that in this classical sort of rigid coupling, the systems A and B lose their independence entirely and as such cease to exist as individual entities; now only the system C exists.

In the case of the argumentally coupling of oscillators, however, the component systems A and B maintain a certain degree of independence and retain very nearly their previous, characteristic oscillatory parameters within the compound system. Hence the result is 3 systems, rather than just one as in the classical case. We might write this as follows:

A + B = { A, B, C }

We should not forget, however, that the argumental coupling between A and B, upon which the integrity and stability of the third system C depends, also constitutes a physical object in its own right. Accordingly, it will be more correct to write:

A + B = { A, B, C, K }

where K denotes the argumental coupling.

In actuality, argumentally coupled oscillators such as Doubochinski's pendulum generally possess not just one, but an entire discrete array of quantized stable modes. Each of these, when realized, constitutes a distinct physical entity with its own characteristic parameters, and each of these is "virtually" present, as a potentiality, in any given coupled state of A and B. We might therefore write, more correctly:

A + B = { A, B, {$C_n$, $K_n$} $_{n = 1,2,3...N}$ }

Later in this article we shall further expand this "nonlinear arithmetic", which we believe expresses a real pathway of development in Nature.

The immensity of the scope of possibilities for the generation of new physical entities, as the offspring of interaction of only two, emerges most clearly, when we add a *second mechanism of interaction of oscillating systems*. discovered together with the initial work on argumental oscillations, as follows.

## 2. Multiply-coupled oscillators and the grouping of resonators in stable configurations

An important subject of radio technology and electrical engineering, is the behavior of networks of coupled resonators, the simplest of these being the famous LCR circuit (Figure 3). Classical theory considers *three basic forms of coupling* between such circuits: inductive coupling, resistive coupling, and capacitive coupling. Figure 4 illustrates the familiar case of inductive coupling, in which a change in current in circuit A induces an electromotive force in circuit B, and visa versa. The result of coupling two independent LCR circuits to each other in this manner, is to produce what classical theory describes as "a single resonator with two degrees of freedom." The resulting system has two proper frequencies, which (in general) are quite different from the proper frequencies of the two uncoupled LCR circuits. In this sense we can say that the original oscillators have "disappeared" or fused together; *they no longer exist as distinct entities within the combined system.* In their place, two oscillation modes appear in the coupled circuit, each of which involves both LCR circuits. The disappearance of the original resonators is underlined by the common electrical engineering practice, of replacing coupled resonators by *equivalent circuits*.

Similar results are obtained for resistive and capacitive coupling. It is emphasized in the classical treatment, that the coupling of resonators in this way can lead to circuits with much broader resonance bands and other properties which are used in the design of filters and other technical devices.

This classical approach has two very crucial limitations, however.

Firstly, the classical treatment never considered, in a systematic way, the *effect of combinations of the three basic forms of coupling.* Thus, in addition to inductive (L), resistive (R) and capacitive (C) couplings, we must consider LR, LC, RC and LCR couplings, each with its own characteristic possibilities. It is found that in the case of combined couplings, the properties of the coupled system *can differ even much more strongly from those of the original, uncoupled resonators, than in the classical, singly-coupled case* [4,5,6,7,10]. For example, in the region of one of its resonant frequencies an LR-coupled circuit (Figure 5) can have a much higher effective Q-value, than either of the component resonators at their resonant frequencies [6]. Moreover, the characteristics of the coupled circuit can be extremely sensitive to changes in the coupling coefficients [6].

Without entering into the details of the analysis, we simply note, that from the point of view of classical mathematical physics, the transition from free, uncoupled resonators to multiply-coupled resonators is equivalent (for free oscillations) to replacing the equation system:

$L_1 X_1'' + R_1 X_1' + C_1 X_1 = 0$
$L_2 X_2'' + R_2 X_2' + C_2 X_2 = 0$

by the equation system for multiple-coupled oscillators:

$L_{11} X_1'' + R_{11} X_1' + C_{11} X_1 + L_{12} X_2'' + R_{12} X_2' + C_{12} X_2 = 0$
$L_{21} X_1'' + R_{21} X_1' + C_{21} X_1 + L_{22} X_2'' + R_{22} X_2' + C_{22} X_2 = 0$

where $L_{12}$, $R_{12}$, $C_{12}$, $L_{21}$, $R_{21}$, $C_{21}$ are coefficients of inductive, resistive and capacitative couplings. In the general case, the zeros on right side must be replaced by functions of time, representing external signals or sources included in the circuits. Since the coupling coefficients enter into the expressions for the coefficients of the fourth degree algebraic equation determining the characteristic values of the differential equation system, it is not surprising that changes in their values can have a significant effect on the proper frequencies, damping and equivalent Q-values of the coupled system.

The physical importance and nearly unlimited possibilities of multiple couplings first become clear, however, when we take into account an additional observation made by the Doubochinskis: *Since the couplings between oscillators carry flows of energy, the physical components involved in those couplings, are subjected to mechanical forces, which in the case of oscillators that are free to move in space, lead to complicated self-organizing motions.*

Suppose, for example, the inductive coupling of two LCR-circuits is achieved by placing the inductive elements of the two circuits (in the form of coils) parallel and near to each other. In that case the coils will experience a varying mechanical force proportional to the product $J_1 \times J_2$ of the currents in the two loops (Figure 6). Assuming the currents are both sinusoidal oscillations with a common frequency f, it is easy to see that the net mechanical force, integrated over a single period, will be proportional to the cosine of the angular phase difference between the two currents. [10]

In most cases of radio engineering, the coils or antennae involved in inductive couplings of oscillating circuits are treated as essentially fixed relative to each other. As a result the mechanical forces generated between them, are ignored in classical analyses of the electrical behavior of the system.

But now imagine that the resonators are able to move freely in space, as indicated in Figure 7. In that case the forces between the inductive elements give rise to accelerations, causing the resonators to change their relative positions. Any change in the relative position of the resonators, in turn, changes the value of the coefficient of mutual induction, and thereby also the oscillatory chacteristics of the coupled system

What we have said about inductive coupling, holds also true for resistive and capacitative forms of coupling. The "feedback" between mechanical motion and electrical oscillations, via variations in the coefficients of coupling and of the relative *phases* of oscillations in the interacting circuits, opens up the possibility of emergence of new forms of combined electro-mechanical oscillations and self-regulating, self-organizing behavior, which have no equivalent in classical treatments of coupled oscillating systems.

Theoretical and experimental investigations carried out by the Doubochinskis and their collaborators at the Vladimir State Pedagogical Institute [4,8,9,10], demonstrated that the simultaneous presence of *more than one dynamic form of coupling* -- for example, combined resistive and inductive couplings – radically transforms the behavior of the coupled system, and leads under certain conditions to a pronounced tendency for coupled oscillating systems to group together in stable formations.

Figure 8 depicts the simplest type of experimental demonstration, in schematic form. Systems $S_1$ and $S_2$ are LCR circuits, where $S_1$ is provided with a sinusoidal voltage source E of frequency f, and $S_2$ operates as a passive resonator. The two circuits are coupled to each other by inductive, capacitative and resistive couplings. Assume further that $S_1$ is fixed, and $S_2$ is free to move with respect to it along the x-axis. Under certain general assumptions on the dependence of the coefficients of coupling on position, each value of the frequency f corresponds to a certain definite separation distance between $S_2$ and $S_1$, at which the net mechanical forces, evoked by the couplings between them, become zero. The system $S_2$ will thus fluctuate around the equilibrium position defined by that separation distance. When the frequency f is changed, $S_2$ moves to occupy the corresponding, new region of stable equilibrium. In general, the equilibrium position is a complicated, piecewise continuous function of the frequency, undergoing discontinuous jumps at certain critical values of f.

In a modified form of this experiment, $S_1$ and $S_2$ are passive resonators, coupled with each other by inductive, resistive and capacitative couplings, and interacting with a third, active oscillating system -- for example a solenoid with a periodic voltage source, or a field of electromagnetic radiation (Figure 9).

It can easily be demonstrated that *phase fluctuations* are essential to the mechanism of self-organization of coupled resonators and to the maintenance of stable constellations formed by them. *Thus, phase fluctuations constitute a common element in both the argumental pendulum, and the tendency of oscillators to group together in stable formations.*

Moreover, under certain conditions it is possible to excite undampened stable oscillations of the resonators around their equilibrium positions, in which the frequency of the spatial (mechanical) oscillation can differ by one or two orders of magnitude from that of the electromagnetic oscillations in the circuits. *This effect combines the principle of dynamic coupling and stable grouping of oscillators, with the principle of argumental oscillations, presented in the previous section of this paper. It points to the existence of a "Macroscopic Quantum Effect for Systems of Coupled Oscillators",* whose implications we shall examine in the following sections. The exchange of energy between the electromagnetic and mechanical oscillations is governed by *phase-frequency modulation*, as in the case Doubochinski's pendulum, but with the difference, that the effect of the mechanical oscillation on the electromagnetic oscillations of the interacting systems must now be treated on an equal footing with the effect of the electromagnetic coupling on their mechanical oscillation.

In the general case, with more than two oscillators and resonators freely moving in space, and coupled by various combinations of inductive, resistive and capacitive couplings, extremely complicated motions are possible, with "phase changes" at critical values of the frequencies of the energy source or sources included among the oscillators. Taking into account both the electrical oscillations and the oscillations of position among the interacting circuits, already three interacting LCR circuits are sufficient to generate an enormous complexity of potential quantized oscillatory modes.

As we noted above in connection with the argumental pendulum, each of those modes constitutes a multiply-coupled oscillating system with its own specific parameters, and therefore deserves to be considered as a distinct physical object in its own right.

The result is, that the combination of the two principles of interaction, described above, makes possible the successive generation of a *practically unlimited number of new physical objects*, starting from the interaction of just a few. At each stage, the already-created objects conserve their identity and individuality, while interacting with each other and external systems to give birth to new objects. This is an essential feature of the "new arithmetic" of interaction, which has no equivalent in the classical Newtonian case of coupled systems.

## 3. A new conception of physical interaction and the generation of physical objects

We now propose to examine the deeper implications of combining the principle of argumental oscillations and the MQE, with principle of aggregation of multiply-coupled oscillators in stable configurations, presented in Sections 1 and 2 respectively. To bring these two phenomena together under a unified conception, it is first necessary to adopt a specific standpoint concerning the character of a real physical object and a true interaction between physical objects (systems), which is neither that of classical physics nor of quantum physics as usually conceived.

Following the tradition of Newton, classical physics tends to think of a *physical object* in a static way, as an entity existing in and of itself, apart from the rest of the Universe.

Instead, we conceptualize a physical object in terms of the *processes of interaction* by which it constitutes itself and maintains its stable existence. A physical object is thus inseparable from its "regime of functioning", and from its interactions with the environment, without which it could not exist. A "physical object" thus signifies a physical system or state or process within a physical system, which exists by virtue of interactions with other physical systems, through which it maintains a certain regime of functioning and certain definite characteristics and parameters, distinguishing it from other physical objects and allowing it, within certain limits, to adapt to external influences without losing its integrity and "identity". Physical objects are by their very nature oscillatory systems.

Our concept of *interaction* also differs substantially from that of classical as well as conventional quantum theory. A *true interaction* presupposes the establishment of a self-regulating regime of exchange of energy between the interacting systems, such that each constantly accommodates and adjusts its internal dynamics to the dynamics of the other, *without either of them losing its essential identity*. A true physical interaction, in our conception, could never exist in a world governed by forces of a strictly Newtonian type, where it is supposed that objects can act upon each other in a purely "abstract" and rigid manner, without any process of adaptation and accommodation, and without the objects themselves being modified internally in any way.

The argumental pendulum, operating in any one of its stable amplitude modes, provides the best presently available illustration of this notion of interaction. As we emphasized in Section 1, the interaction between the mechanical pendulum and the electromagnet involves a self-regulating regime of exchange of energy between the two systems, whose essential mechanism is phase fluctuations. That self-regulating process of exchange therefore constitutes not only a *true physical interaction*, but at the same time also a *definite physical object* in the sense defined above. Thus, not one, but *two* new physical objects arise in each quantized state of Doubochinski's pendulum:

> (1) the *coupled system* of pendulum and electromagnet, operating at a given stable amplitude; and
> (2) the *regime of interaction* providing the self-regulating exchange of energy, upon which the given mode is based.

It is important to stress, that although the *coupled system* and the *dynamic coupling*, *corresponding to a given regime of interaction,* are in a sense facets of one and the same process -- neither of which could exist without the other --, they nevertheless must be regarded as *distinct* physical objects. The regime of the coupled system is characterized by the amplitude and period of the pendulum oscillation, while the regime of the interaction is characterized by the fluctuations in phase relations between the electromagnet and pendulum.

Similar observations hold for the stable configurations of two or more multiply-coupled resonators, with the important further difference, that each of the different types of couplings -- inductive, resistive and capacitive -- constitutes a physically distinct channel of interaction. This in turn leads to a much greater proliferation of potential physical objects.

## 4. The electromagneto-mechanical dual nature of physical interactions

Both of the model systems considered here – the Doubochinski pendulum and aggregates of multiply-coupled LCR resonators in a high-frequency field – constitute electromechanical (or better, electromagneto-mechanical) systems, whose essential characteristics derive from a certain interlinking of electromagnetic oscillations with periodic motions in space. The symbiosis of electromagnetic and mechanical oscillations lies at the heart of the Macroscopic Quantum Effect and the aggregative tendency of oscillators, described above. Reflecting upon the significance of this fact, we have been led to some far-reaching conclusions concerning the nature of real physical interactions in general. The following remarks are intended only as a "zeroth approximation" to indicate the general direction of our thoughts on this topic, leaving more thorough discussion to a future occasion.

1. Real physical objects -- as opposed to abstractions invented for the purposes of mathematical modelling – are invariably oscillatory in nature, and their existence is invariably connected with oscillatory motion in space as well as oscillations of an electromagnetic character. The maintenance and stability of each object depends on the "live", fluctuating character of its interactions.

2. Classical electrodynamics describes the interrelationship between mechanical and electromagnetic oscillations in terms of electromotive and ponderomotive effects. We prefer to speak of the *electromagneto-mechanical dual nature of physical processes*, in which the mechanical and electromagnetic components are organically linked to each other as facets of a single unity, in a manner analogous to the real and imaginary parts of a complex analytic function.[2]

3. To the extent physical objects constitute electromagneto-mechanical oscillating systems, their interactions are intrinsically argumental in character, leading to phase-frequency modulation and to conditions under which the quantization and grouping effects, discussed above, are likely to occur. A simple heuristic model of this would be a pair of freely-moving high-frequency LCR-oscillators whose electromagnetic interaction is modulated by, and thereby argumentally coupled to, a lower-frequency quasi-periodic relative motion of the systems in space.[3]

4. It is typical of the economy of Nature, that the principle of argumental interaction of electromagneto-mechanical oscillatory systems, is sufficient to generate a practically unlimited wealth of physical objects from the interaction of just a few, without the need for additional assumptions.

---

[2] This approach may prove fruitful in conceptualizing many electrodynamic phenomena, including for example the phenomenon of radiation pressure, for which the standard treatments can hardly be considered satisfactory.

[3] We must not forget, of course, that such systems never exist in isolation, but owe their existence and functioning to their interaction with the entire environment.

*Bibliography*

1. M. JUNG et al. *First Observation of Bound-state Beta-Decay,* Physical Review Letters*,* Vol. 69, No. 15, pp. 2164-2167, 1992; F. BOSCH et al., *Observation of Bound-state Beta-Decay of Fully Ionized 187Re*, Physical Review Letters*,* Vol. 77, No. 26, pp. 5190-5193, 1996; P. KIENLE, *Beta-decay Experiments and Astrophysical Implications* in N. Prantzos and S. Harissopulus, *Proceedings: Nuclei in the Cosmos*, pp.181-186, 1999.

2. D.B. DOUBOCHINSKI, J. TENNENBAUM, *The Macroscopic Quantum Effect in Nonlinear Oscillating Systems: a Possible Bridge between Classical and Quantum Physics*, paper delivered to Moscow seminar on "Atomic Structure, New ideas and Perspectives", January 2007. http://arxiv.org/pdf/0711.4892

3. J. TENNENBAUM *Amplitude Quantization as an Elementary Property of Macroscopic Vibrating Systems.* 21st CENTURY SCIENCE & TECHNOLOGY - USA Winter 2005 – 2006.

4. D.B. DOUBOCHINSKI, Ya.B. DOUBOCHINSKY et al., *Electromagnetic model of the interaction of resonators.* Dokl. Akad. Nauk SSSR, 227, N°3, 596 (1976) [Sov.Phys.Doklady 27, 51(1976)].

5. Ya.B. DOUBOCHINSKI, *On certain properties of systems of resonators*, (in Russian). Uchenye Zapiski, Vladimir State Pedagogical Institute, Vladimir (1971).

6. M.I. KOZAKOV, *Investigations of coupled resonators* (in Russian). Uchenye Zapiski, Vladimir State Pedagogical Institute, Vladimir (1971).

7. D. I. PENNER, Ya.B. DOUBOCHINSKI, *Assimilative potential of coupled oscillating systems* (in Russian). Uchenye Zapiski, Vladimir State Pedagogical Institute, Vladimir (1972).

8. D.B. DOUBOCHINSKI, Ya.B. DOUBOCHINSKI, D.I. PENNER, *On the forces between resonators, coupled by a single dynamic coupling* (in Russian). Uchenye Zapiski, Vladimir State Pedagogical Institute, Vladimir (1972).

9. Ya. B. DOUBOCHINSKI, *On an electromechanical model system* (in Russian), Uchenye Zapiski, Vladimir State Pedagogical Institute, Vladimir (1972).

10. D.I. PENNER, Ya.B. DOUBOCHINSKI, D.B. DOUBOCHINSKI, *On certain aspects of the problem of interacting resonators* (in Russian), Uchenye Zapiski, Vladimir State Pedagogical Institute, Vladimir (1974).

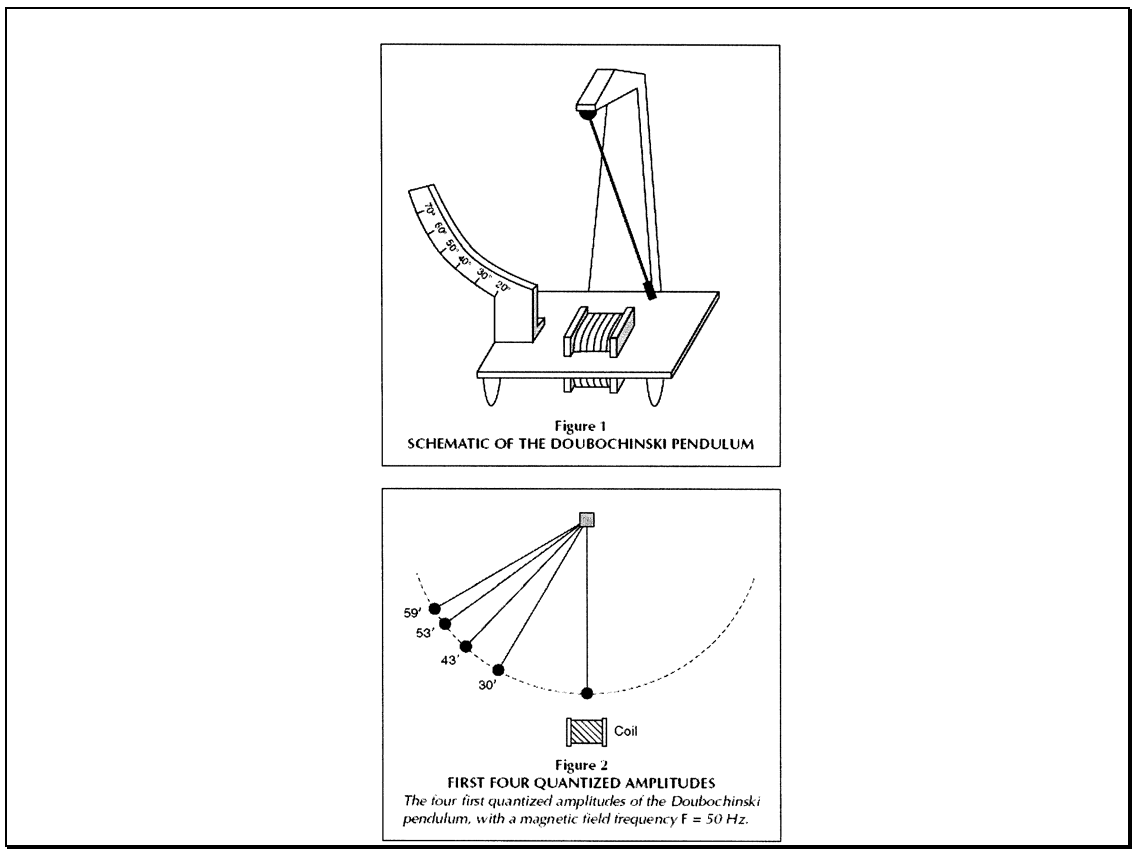

**Figure 1**
SCHEMATIC OF THE DOUBOCHINSKI PENDULUM

**Figure 2**
FIRST FOUR QUANTIZED AMPLITUDES
*The four first quantized amplitudes of the Doubochinski pendulum, with a magnetic field frequency F = 50 Hz.*

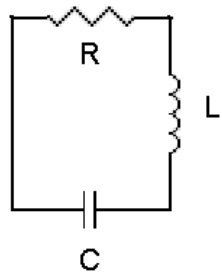

Figure 3  Simple LCR circuit

Figure 4  Inductively coupled oscillators according to classical physics

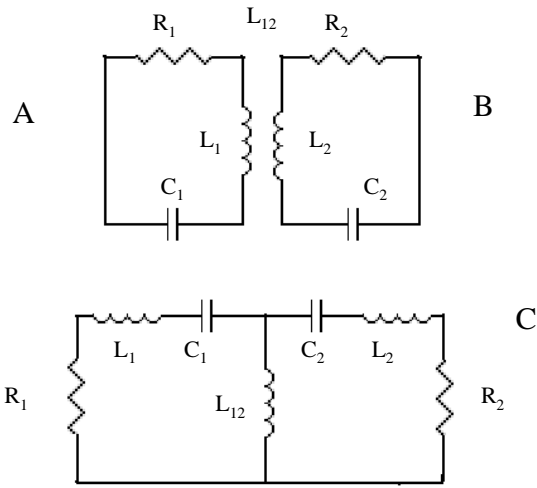

"Equivalent circuit": A and B "disappear" into new system C

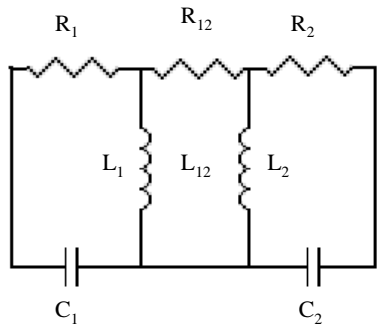

Figure 5 LR-coupled circuit

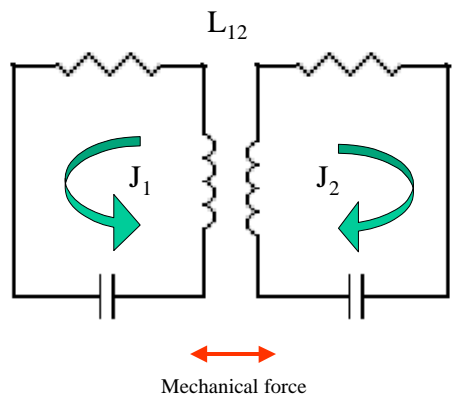

Figure 6 Force between inductively coupled circuits proportional to $J_1 \times J_2$

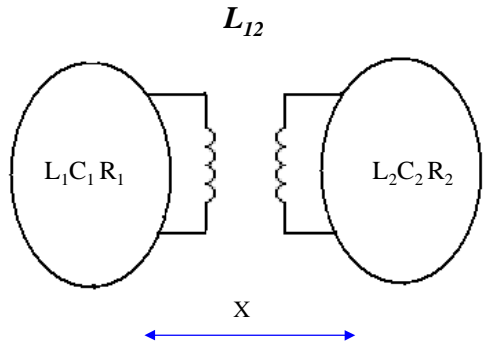

Figure 7  Interaction of moveable oscillators

Resonators free to move in X-direction. The coefficient of inductive coupling $L_{12}$ is a function of relative position along the X-axis

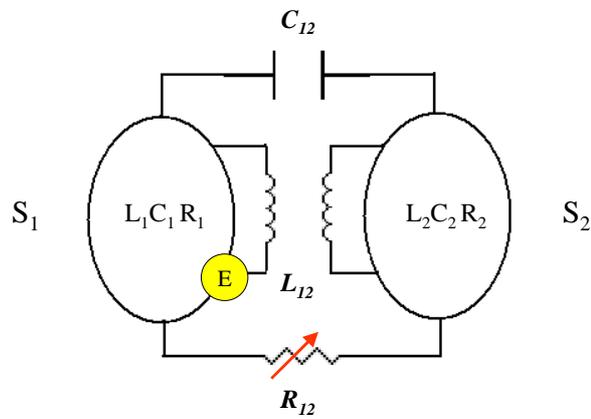

Figure 8  Interaction of triply-coupled $\{L_{12}\ C_{12}\ R_{12}\}$, free to move in all three space directions (not shown in drawing). Each degree of freedom of motion "modulates" one of coefficients of coupling.

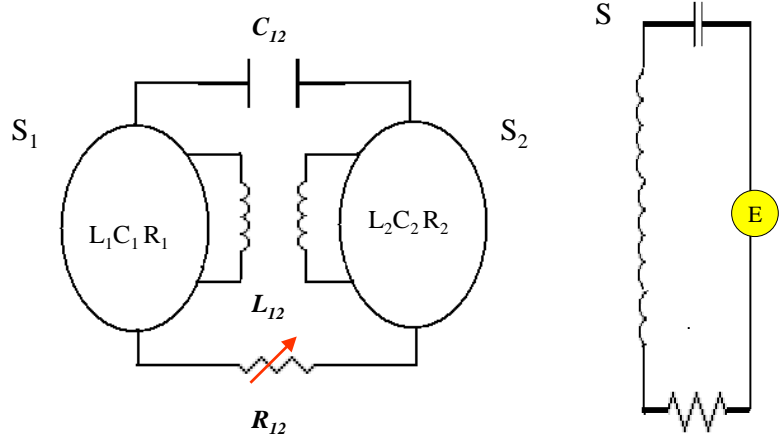

Figure 9 Triply-coupled {$L_{12}\,C_{12}\,R_{12}$} interaction of two oscillators in field of solenoid. The oscillators can move freely in all three space directions, one for each type of coupling, thereby modulating the corresponding coupling coefficients